\documentclass[aps,prd,twocolumn,superscriptaddress,showpacs,showkeys,floatfix]{revtex4-1}

\RequirePackage{graphicx}
\RequirePackage{bm}
\RequirePackage{amssymb}

\begin{document}

\title{The shadow of dark matter as a shadow of string theory}



\author{Andrii Dashko}
\email{dashko.andriy@gmail.com}
\affiliation{Department of Physics and Engineering Physics, 
University of Saskatchewan, 116 Science Place, Saskatoon, Canada SK S7N 5E2}
\affiliation{Department of Physics,
Taras Shevchenko National University of Kyiv, 64/13, Volodymyrska Street, Kyiv 01601, Ukraine}

\author{Rainer Dick}
\email{rainer.dick@usask.ca}

\affiliation{Department of Physics and Engineering Physics, 
University of Saskatchewan, 116 Science Place, Saskatoon, Canada SK S7N 5E2}



\begin{abstract}  
We point out that string theory can solve the conundrum to explain the emergence
of an elec\-tro\-weak dipole moment from elec\-tro\-weak singlets through induction of those
dipole moments through a Kalb-Ramond dipole coupling. 
This can generate a $U_Y(1)$ portal to dark matter and entails the possibility
that the $U_Y(1)$ gauge field is related to a fundamental vector field for open 
string interactions.
The requirement to explain the observed dark matter abundance relates the coupling scale
$M$ in the corresponding low-energy effective $U_Y(1)$ portal to the dark matter mass $m_\chi$.
The corresponding electron recoil cross sections for a single dipole coupled dark matter
species are generically below the current limits from XENON, SuperCDMS and SENSEI, except 
in the GeV mass range if the electric dipole coupling becomes stronger than the magnetic 
coupling, $a_e^2\ge a_m^2$. Furthermore, the recoil cross sections are above the 
neutrino floor and the $U_Y(1)$ portal can be tested with longer exposure or larger 
detectors. Discovery of electroweak dipole dark matter would therefore open
an interesting window into string phenomenology.
\end{abstract}


\maketitle


\section{Introduction\label{sec:intro}}

The puzzle of the large dark matter densities in galaxies and galaxy clusters remains an 
enigma for particle physics. The fact that a hitherto unobserved particle with weak strength
couplings to Standard Model particles can generate the observed dark matter abundance through
thermal freeze-out from the primordial heat bath (the ``WIMP miracle'') continues to nourish
hopes that dark matter may not only reveal itself through gravitational interactions, but can 
also be detected in particle physics labs \cite{ed}. Following the designation of 
the \textit{Higgs portal} \cite{frank} for dark matter models where the interaction is mediated 
by Higgs exchange \cite{zee,mcdonald,bento,cliff,dklm,wells,maxim,yann} 
(see \cite{rd1,gambit} for recent reviews), 
the notion of ``portals'' for the non-gravitational interaction between dark matter and the 
Standard Model has been widely adopted, including neutrino portals and vector portals. 
These standard options for non-gravitational dark matter couplings usually do not include a 
photon portal, as the optical darkness of the dominant matter component in large scale astronomical 
structures is usually assumed to be a consequence of the absence of direct photon couplings. 
However, Sigurdson \textit{et al.} had
pointed out that dipole couplings of MeV or GeV scale dark matter to photons comply with the 
darkness requirement if the coupling is sufficiently suppressed \cite{kris1}, see 
also \cite{masso,barger,heo1,tait,cline,heo2},
and Profumo and Sigurdson coined the notion of a ``shadow of dark matter'' for 
this scenario \cite{PS}. Possible dipole couplings to photons involve
dark fermions in the form 
\begin{equation} \label{eq:PS}
\mathcal{L}=\frac{1}{2M_d}F^{\mu\nu}\overline{\chi}_1S_{\mu\nu}(a_m+\mathrm{i}a_e\gamma_5)\chi_2
+\mathrm{h.c.}, 
\end{equation}
where here we use $S_{\mu\nu}=\mathrm{i}[\gamma_\mu,\gamma_\nu]/4=\gamma^0 S^+_{\mu\nu}\gamma^0$.
The terms in Eq.~(\ref{eq:PS}) yield magnetic and electric Pauli terms
\begin{equation}\label{eq:PS2}
\mathcal{L}\to\frac{1}{2M_d}\psi_1^+(a_m\bm{B}+a_e\bm{E})\cdot\bm{\sigma}\psi_2+\mathrm{h.c.}
\end{equation}
in the non-relativistic limit.

Couplings of the form (\ref{eq:PS}) were also used in \cite{masso,also1,banks,also2,also3} 
in proposals to explain the DAMA annual modulation signal in nuclear recoils.
More recently, Conlon \textit{et al.} pointed out that the direct photon coupling proposed by
Profumo and Sigurdson can reconcile the 3.5 keV data from the Hitomi, XMM-Newton and Chandra
observations of the Perseus cluster through X-ray absorption and resubmission \cite{conlon}.

Of course, a mass-suppressed photon portal \textit{per se} to electro\-weak singlet dark matter
breaks the electro\-weak symmetry of the Standard Model. Therefore it makes sense to replace the 
mass suppressed photon portal with a mass suppressed $B_{\mu\nu}$ portal (or $U_Y(1)$ portal) 
where $B_{\mu\nu}=\partial_\mu B_\nu-\partial_\nu B_\mu$ is the field strength of the 
electroweak $U_Y(1)$ symmetry. This then automatically entails the photon portal through
electroweak mixing into mass eigenstates, $B_\mu=A_\mu\cos\theta-Z_\mu\sin\theta$ and
leaves the electroweak symmetries unbroken. Indeed, it was noticed already by Cline, Moore 
and Frey \cite{cline} that couplings of the form (\ref{eq:PS}) should also entail corresponding 
couplings to the $Z$ boson.

We wish to draw attention to the fact that the Kalb-Ramond field of string theory can help 
to generate dipole couplings of the form (\ref{eq:PS}). The Kalb-Ramond field is an 
anti-symmetric tensor field $C=C_{\mu\nu}dx^\mu\wedge dx^\nu/2$ which does not need to be 
closed, $dC\neq 0$, and therefore cannot simply be considered as the field strength of a 
hidden $U(1)$ symmetry. 

It has recently been pointed out that the strongest constraints for low mass dipole
coupled dark matter should arise from direct searches in electron recoils \cite{rouven1}.
Therefore we also discuss the corresponding electron 
recoil constraints under the assumption of generation from thermal freeze-out.

The natural emergence of couplings of the Kalb-Ramond field to $U(1)$ gauge fields 
is reviewed in Sec. \ref{sec:strings}. The ensuing possibility that the Kalb-Ramond field 
can induce electroweak dipole couplings for electroweak singlets is introduced in 
Sec. \ref{sec:dipoles}.
Abundance constraints on the magnetic dipole coupling scale $M^{-1}=a_m/M_d$ for a single dipole 
coupled dark matter species $\chi$ and the resulting constraints from direct dark matter searches
in electron recoils are discussed in Secs. \ref{sec:M} and \ref{sec:recoil}, respectively.
Section \ref{sec:conc} summarizes our conclusions.

\section{A shadow of string theory\label{sec:strings}}

Closed strings contain anti-symmetric tensor excitations in their low-energy 
sector through the anti-symmetric Lorentz-irreducible component of the oscillation
states $(a_{+,1}^\mu)^+ (a_{-,1}^\nu)^+|0\rangle$ \cite{GSW,BLT}. Anti-symmetric tensor fields can
also mediate gauge interactions between string world sheets \cite{KR0}, and these fields also
participate in brane interactions \cite{KR1,KR2,KR3,KR4,KR5,KR6}. 

There are two ways how the Kalb-Ramond field can couple to $U(1)$ gauge bosons, and both 
of them are related to the gauge symmetries of string-string interactions. 
We therefore need to review the string couplings of the Kalb-Ramond field and how 
they necessitate a coupling to $U(1)$ gauge bosons in the presence of open strings. Kalb 
and Ramond had generalized the work of Feynman and Wheeler for action at a distance in 
electrodynamics in their seminal work, but with the wisdom of hindsight it is easier
to start with the Lagrangian formulation of the pertinent string couplings. 
This formulation also shows how to generalize the construction for couplings to
several $U(1)$ gauge fields, and demonstrates that we can keep the $U(1)$ gauge fields for 
the boundary charges of open strings massless.

Gauge interactions between strings can be described in analogy to
electromagnetic interactions if the basic Nambu-Goto action is amended
with a coupling term to the Kalb-Ramond field $C_{\mu\nu}=-C_{\nu\mu}$ \cite{KR0}
(we avoid the usual designation $B_{\mu\nu}$ for the Kalb-Ramond field to avoid confusion 
with the $U_Y(1)$ field strength tensor),
\begin{eqnarray}\nonumber
S&=&-\sum_a T\int\! d\tau_a\int_0^\ell\! d\sigma_a
\sqrt{(\dot{x}_a\cdot x_a')^2-\dot{x}_a^2x'^2_a}
\\ \nonumber
&&+\sum_a\frac{\mu_s}{2}\int\! d\tau_a\int_0^\ell\! d\sigma_a\left(\dot{x}_a^\mu x'^\nu_a
-\dot{x}_a^\nu x'^\mu_a\right)C_{\mu\nu}(x_a)
\\ \label{eq:action1}
&&+\sum_a g\!\int\! d\tau_a\left[\dot{x}_a^\mu \mathcal{B}_{\mu}(x_a)\right]_{\sigma_a=0}^{\sigma_a=\ell}.
\end{eqnarray}
Here $T$ is the string tension, $\mu_s$ is a string coupling constant (or \textit{string charge}) 
with the dimension of mass, $\tau_a$
and $\sigma_a$ are timelike and spacelike coordinates on the world sheet of the $a$-th string, 
respectively, and $x_a\equiv x(\tau_a,\sigma_a)$ describes the embedding of the string world 
sheet into spacetime. The world sheet string interaction term can be written in 
the form $\mu_s\int C$, just like the electromagnetic interaction term in particle 
physics for particles of charge $q$ can be written as a world line integral $q\int A$.
The dimensionless charge $g$ appears only on the endpoints of open strings.

To appreciate the connection of the action (\ref{eq:action1}) to the emergence of gauge 
dipoles from gauge singlet fields, we need to consider the resulting string equations of motion.
 For simplicity of the left hand sides, we display these equations in flat Minkowski spacetime in 
conformal gauge $\dot{x}_a^2+x'^2_a=0$, $\dot{x}_a\cdot x_a'=0$ (see Ref.~\cite{conf} for a general 
proof of existence of conformal gauge in Minkowski signature). The world sheet equation of motion is
\begin{equation}\label{eq:stringbulk}
2T\left(\ddot{x}_{a\mu}-x''_{a\mu}\right)
=\mu_s C_{\mu\nu\rho}(x_a)\left(\dot{x}_a^\nu x'^\rho_a
-\dot{x}_a^\rho x'^\nu_a\right),
\end{equation}
where 
\begin{equation}\label{eq:C3}
C_{\mu\nu\rho}=\partial_\mu C_{\nu\rho}+\partial_\nu C_{\rho\mu}+\partial_\rho C_{\mu\nu}
\end{equation} 
are the components of the 3-form field strength of the Kalb-Ramond field.
We will also write this in the short form $C_3=dC_2$.

The additional conditions at the boundaries of open strings are
with $\mathcal{B}_{\mu\nu}=\partial_\mu\mathcal{B}_\nu-\partial_\nu\mathcal{B}_\mu$,
\begin{equation}\label{eq:boundaries}
\Big[T x'_{a\mu}+\left[g\mathcal{B}_{\mu\nu}(x_a)
-\mu_s C_{\mu\nu}(x_a)\right]\dot{x}_a^\nu\Big]_{\sigma\in\{0,\ell\}}=0.
\end{equation}
The string equations of motion (\ref{eq:stringbulk},\ref{eq:boundaries})
are invariant under the \textit{KR gauge symmetry} 
\begin{equation}\label{eq:KR1}
C_{\mu\nu}\to C_{\mu\nu}+\partial_\mu f_\nu-\partial_\nu f_\mu,\quad
\mathcal{B}_\mu\to\mathcal{B}_\mu+(\mu_s/g)f_\mu,
\end{equation}
and under the $U(1)$ gauge transformation $\mathcal{B}_\mu\to\mathcal{B}_\mu +\partial_\mu f$.

The coupling terms in the action (\ref{eq:action1}) imply that strings are sources of the 
Kalb-Ramond field $C$ and the accompanying vector field $\mathcal{B}$, and the action should 
be amended with kinetic terms for those fields. The KR gauge symmetry (\ref{eq:KR1}) is 
preserved through the kinetic term
\begin{eqnarray}\nonumber
\mathcal{L}&=&-\,\frac{1}{6}C^{\mu\nu\rho}C_{\mu\nu\rho}-\frac{1}{4}\mathcal{B}^{\mu\nu}\mathcal{B}_{\mu\nu}
+\frac{\mu_s}{2g}C^{\mu\nu}\mathcal{B}_{\mu\nu}
\\ \label{eq:kinfields}
&&-\,\frac{\mu_s^2}{4g^2}C^{\mu\nu}C_{\mu\nu}.
\end{eqnarray}
 For the normalization of the kinetic term for the Kalb-Ramond field, we note that its variation 
under variation $\delta C_{\mu\nu}$ of the Kalb-Ramond field is
\begin{eqnarray}\nonumber
\delta(C^{\mu\nu\rho}C_{\mu\nu\rho})&=&2C^{\mu\nu\rho}\delta C_{\mu\nu\rho}
\\ \nonumber
&=&2C^{\mu\nu\rho}(\partial_\mu\delta C_{\nu\rho}+\partial_\nu\delta C_{\rho\mu}+\partial_\rho\delta C_{\mu\nu})
\\ \label{eq:vari}
&=&6C^{\mu\nu\rho}\partial_\mu\delta C_{\nu\rho}.
\end{eqnarray}
The field equations are then
\begin{equation}\label{eq:sources1}
\partial_\mu C^{\mu\nu\rho}(x)+\frac{\mu_s}{2g}\mathcal{B}^{\nu\rho}(x)
-\frac{\mu_s^2}{2g^2} C^{\nu\rho}(x)=-\sum_a j_a^{\nu\rho}(x)
\end{equation}
and
\begin{equation}\label{eq:sources2}
\partial_\mu\left(\mathcal{B}^{\mu\nu}(x)-\frac{\mu_s}{g}C^{\mu\nu}(x)\right)=-\sum_a j_a^{\nu}(x),
\end{equation}
where from Eq.~(\ref{eq:action1}) the string charge currents are
\begin{eqnarray}\nonumber
j_a^{\mu\nu}(x)&=&\frac{\mu_s}{2}\int\! d\tau_a\!\int_0^\ell\! d\sigma_a
\left[\dot{y}^\mu(\tau_a,\sigma_a) y'^\nu(\tau_a,\sigma_a)
\right.
\\ \label{eq:stringcurrents1}
&&-\left.\dot{y}^\nu(\tau_a,\sigma_a) y'^\mu(\tau_a,\sigma_a)\right]\delta(x-y(\tau_a,\sigma_a))
\end{eqnarray}
and
\begin{equation}\label{eq:stringcurrents2}
j_a^{\mu}(x)=g\!\int\! d\tau_a\left[\dot{y}^\mu(\tau_a,\sigma_a)
\delta(x-y(\tau_a,\sigma_a))\right]_{\sigma_a=0}^{\sigma_a=\ell}.
\end{equation}
The boundary current $j_a^{\mu}(x)$ is the combination of $U(1)$ currents of a charge $g$
at $\sigma_a=\ell$ and a charge $-g$ at $\sigma_a=0$. Up to boundary terms 
at $\tau_a\to\pm\infty$ (which also appear in the currents of charged particles
in electrodynamics), the currents satisfy $\partial_\mu j_a^{\mu\nu}(x)=(\mu_s/2g)j_a^\nu(x)$ 
and $\partial_\mu j_a^{\mu}(x)=0$ \cite{KR0}.

A very important lesson from the work of Kalb and Ramond is
that the world sheet gauge field $C_{\mu\nu}$ in the presence of open strings
has a mass term and couples to $U(1)$ gauge fields $\mathcal{B}_\mu$ in 
the form $C_{\mu\nu}\mathcal{B}^{\mu\nu}$. Indeed, we can easily generalize the
construction to the case of different boundary charges $g_I$ for different types of open
strings with corresponding gauge fields $\mathcal{B}_{I,\mu}$.
We can simply replace the boundary term in Eq.~(\ref{eq:action1}) according to
\begin{eqnarray}\nonumber
&&\sum_a g\!\int\! d\tau_a\left[\dot{x}_a^\mu \mathcal{B}_{\mu}(x_a)\right]_{\sigma_a=0}^{\sigma_a=\ell}
\\ \label{eq:gen1}
&&\to
\sum_a g_{I(a)}\!\int\! d\tau_a\left[\dot{x}_a^\mu \mathcal{B}_{I(a),\mu}(x_a)
\right]_{\sigma_a=0}^{\sigma_a=\ell},
\end{eqnarray}
where $g_{I(a)}$ is the boundary charge of the $a$-th string.
The boundary equation (\ref{eq:boundaries}) for the $a$-th string becomes  
\begin{eqnarray}\nonumber 
&&\Big[T x'_{a\mu}+\left[g_{I(a)}\mathcal{B}_{I(a),\mu\nu}(x_a)
-\mu_s C_{\mu\nu}(x_a)\right]\dot{x}_a^\nu\Big]_{\sigma\in\{0,\ell\}}
\\ \label{eq:boundaries2}
&&\quad =0.
\end{eqnarray}
The $U(1)$ gauge fields transform under KR symmetry according to
$\mathcal{B}_{I,\mu}\to\mathcal{B}_{I,\mu}+(\mu_s/g_I)f_\mu$, and the KR gauge kinetic term
becomes
\begin{eqnarray}\nonumber
\mathcal{L}&=&-\,\frac{1}{6}C^{\mu\nu\rho}C_{\mu\nu\rho}
-\frac{1}{4}\sum_I\mathcal{B}_I^{\mu\nu}\mathcal{B}^I_{\mu\nu}
+\sum_I\frac{\mu_s}{2g_I}C_{\mu\nu}\mathcal{B}_I^{\mu\nu}
\\ \label{eq:kinfields2}
&&-\sum_I\frac{\mu_s^2}{4g_I^2}C^{\mu\nu}C_{\mu\nu}.
\end{eqnarray}
The KR symmetric equation (\ref{eq:sources1}) acquires sums over the open string classes $I$
in the $g_I$-dependent terms on the left hand side, and Eq.~(\ref{eq:sources2}) generalizes
to
\begin{equation}\label{eq:sources2J}
\partial_\mu\left(\mathcal{B}_J^{\mu\nu}(x)-\frac{\mu_s}{g_J}C^{\mu\nu}(x)\right)
=-\sum_{a,I(a)=J} j_a^{\nu}(x),
\end{equation}
where the sum on the right hand side includes only the open strings which carry 
boundary charge $g_J$. Finally, the currents for the $a$-th string satisfy
$\partial_\mu j_a^{\mu\nu}(x)=(\mu_s/2g_{I(a)})j_a^\nu(x)$. 

\section{Electroweak dipoles induced by the Kalb-Ramond field\label{sec:dipoles}}

 From the point of view of string phenomenology, it is interesting to explore 
the possibility that the abelian gauge field $B_\mu$ of the Standard Model can 
also couple to the boundary charges of open strings. This would imply in particular
a coupling $C_{\mu\nu}B^{\mu\nu}$ to the Kalb-Ramond field which is a necessary
ingredient to generate the $U_Y(1)$ portal from a renormalizable model. 
The $U_Y(1)$ portal can arise from an extension of the Standard Model of the form
\begin{eqnarray}\nonumber
\mathcal{L}_{BC\chi}&=&\overline{\chi}\left(\mathrm{i}\gamma^\mu\partial_\mu
-m_\chi\right)\chi-\frac{1}{6}C^{\mu\nu\rho}C_{\mu\nu\rho}
\\  \nonumber
&&-\,\frac{1}{2}m^2_C C_{\mu\nu}C^{\mu\nu}
-g_{BC}m_C B^{\mu\nu}C_{\mu\nu}
\\  \label{eq:L2}
&&-\,g_{C\chi}\overline{\chi}S^{\mu\nu}(a_m+\mathrm{i}a_e\gamma_5)\chi C_{\mu\nu},
\end{eqnarray}
where perturbatively small couplings $g_{BC}$ and $g_{C\chi}$ are assumed,
and the matrices $S_{\mu\nu}=\mathrm{i}[\gamma_\mu,\gamma_\nu]/4$ are the 4-spinor
representations of the Lorentz generators. 
Together with the kinetic term $-B_{\mu\nu}B^{\mu\nu}/4$ for the $U_Y(1)$ gauge field
(not listed in equation (\ref{eq:L2}) as this is already included in the Standard Model),
the bosonic terms in the Lagrangian (\ref{eq:L2}) are KR gauge symmetric 
if $g_{BC}=1/\sqrt{2}$.
However, KR gauge symmetry will likely be broken at low energies in the process of moduli 
stabilization \cite{modh1,modh2,modh3,modh4,modh5,mod1,mod1b,mod2,mod3,mod4,mod5,mod6,mod7,mod8,mod9,mod10}.
The Kalb-Ramond field can contribute to moduli stabilization in particular 
through $H$-type fluxes \cite{modh1,modh2,modh3,modh4,modh5,mod1,mod1b,mod9} through 
the internal components of the gauge invariant 
3-form $C_3=C_{KLM}dx^K\wedge dx^L\wedge dx^M/6$. These fluxes would imply
\begin{equation}
\langle C_3^2\rangle\equiv\sum_{L,M,N=4}^9\langle C_{LMN}C^{LMN}\rangle>0.
\end{equation}
The fluxes and dilaton stabilization, e.g. for $\langle C_3^2\rangle\ll m_\phi^2 f_\phi^2$,
\begin{eqnarray}\nonumber
m^2_\phi\phi-\frac{1}{6f_\phi}\exp(-\phi/f_\phi)\langle C_3^2\rangle
&\simeq& m^2_\phi\phi-\frac{1}{6f_\phi}\langle C_3^2\rangle
\\
&=&0,
\end{eqnarray}
\begin{eqnarray}\nonumber
&&\frac{1}{6}\exp(-\phi/f_\phi)C_3^2
\\
&&\to
\frac{1}{6}C_{\mu\nu\rho}C^{\mu\nu\rho}-\frac{1}{36 m_\phi^2 f_\phi^2}\langle C_3^2\rangle
C_{\mu\nu\rho}C^{\mu\nu\rho},
\end{eqnarray}
would change the normalization of the four-dimensional remnant of the Kalb-Ramond field,
thus changing the ratio of the mass and coupling terms in the low-energy effective action
(\ref{eq:L2}). This  breaks the Kalb-Ramond gauge symmetry and demotes 
the Kalb-Ramond field from an anti-symmetric gauge tensor field to an anti-symmetric matter
tensor field. Magnetic dipole coupling terms are expected in anti-symmetric matter tensor 
theories since it has been shown that they provide consistent renormalizable source 
terms for anti-symmetric matter tensor fields \cite{mtf1,mtf2,mtf3}. Since KR gauge symmetry
is the only symmetry which could prevent their generation, they would seem unavoidable
once the KR gauge symmetry is broken, contrary to the electric dipole moments, which
can be prevented by parity symmetry. Furthermore, will see below that the magnetic dipole 
couplings determine the relic abundance from thermal freeze-out in theories with both 
magnetic and electric dipole couplings. We are therefore primarily interested in the 
constraints on the magnetic dipole coupling, but we also carry through the electric dipole 
coupling for completeness. We also note that the Kalb-Ramond formulation of open string 
interactions treats open strings as elementary $U(1)$ dipoles, and therefore
dipole interaction terms and KR gauge symmetry breaking would appear unavoidable in any
low energy field theory formulation of the theory which would be based on renormalizable
terms. 

Elimination of the Kalb-Ramond field for energies much smaller than $m_C$
yields a $U_Y(1)$ portal for the dark fermions $\chi$ which appears as a $\gamma,Z$ portal
in terms of mass eigenstates,
\begin{eqnarray} \nonumber
\mathcal{L}_{B\chi}&=&\frac{g_{BC}g_{C\chi}}{m_C}B_{\mu\nu}
\overline{\chi}S^{\mu\nu}(a_m+\mathrm{i}a_e\gamma_5)\chi
\\ \nonumber
&=&\frac{g_{BC}g_{C\chi}}{m_C}(F_{\mu\nu}\cos\theta-Z_{\mu\nu}\sin\theta)
\\ \label{eq:ps1}
&&\times\overline{\chi}S^{\mu\nu}(a_m+\mathrm{i}a_e\gamma_5)\chi.
\end{eqnarray}
Here $\theta$ is the weak mixing angle, $B_\mu=A_\mu\cos\theta-Z_\mu\sin\theta$. 
The induction of the coupling term (\ref{eq:ps1}) from equation (\ref{eq:L2}) shows
that the seemingly paradoxical notion of electroweak dipole moments of electroweak
singlets is resolved in string theory through transfer of Kalb-Ramond dipoles 
into the $U_Y(1)$ sector through the Kalb-Ramond field.

The gauge fields $\mathcal{B}_{I,\mu}$ disappear if we only have
closed strings. Invariance of couplings under the remaining KR gauge 
symmetry $C_{\mu\nu}\to C_{\mu\nu}+\partial_\mu f_\nu-\partial_\nu f_\mu$ then
allows for a Cremmer-Scherk coupling $\epsilon^{\mu\nu\rho\sigma}C_{\mu\nu}B_{\rho\sigma}$
between Kalb-Ramond fields and $U(1)$ field strengths \cite{CS}, and integrating this out for 
massive Kalb-Ramond fields can also generate gauge invariant low-energy effective
dipole couplings. Elimination of $C_{\mu\nu}$ from the Lagrangian
\begin{eqnarray}\nonumber
\mathcal{L}_{BC\chi}^{(2)}&=&\overline{\chi}\left(\mathrm{i}\gamma^\mu\partial_\mu
-m_\chi\right)\chi
-\frac{1}{6}C^{\mu\nu\rho}C_{\mu\nu\rho}
\\ \nonumber
&&-\,\frac{1}{2}m^2_C C_{\mu\nu}C^{\mu\nu}
-\frac{1}{2}g_{BC}m_C\epsilon^{\mu\nu\rho\sigma}C_{\mu\nu} B_{\rho\sigma}
\\ \label{eq:L1}
&&-\,\frac{1}{2}
g_{C\chi}\epsilon^{\mu\nu\rho\sigma}C_{\mu\nu}
\overline{\chi}S_{\rho\sigma}(a_m+\mathrm{i}a_e\gamma_5)\chi
\end{eqnarray}
 for energies much smaller than $m_C$ yields again the $U_Y(1)$
portal (\ref{eq:ps1}).

We have formulated the Kalb-Ramond induced generation of dipole coupling terms between 
gauge fields and dark fermions for a single fermion species, but it is clear that with 
the substitution 
\begin{eqnarray}\nonumber
&&g_{C\chi}\overline{\chi}S^{\mu\nu}(a_m+\mathrm{i}a_e\gamma_5)\chi
\\
&&\to\frac{1}{2}\sum_{i,j}g_{Cij}\overline{\chi}_iS^{\mu\nu}(a_m+\mathrm{i}a_e\gamma_5)\chi_j
+\mathrm{h.c.},
\end{eqnarray}
this mechanism works for any number of dark fermion species, and can generate in particular
the coupling which is required for photon absorption by a dark two-level system.

\section{Dark matter abundance constraints on the $U_Y(1)$ portal\label{sec:M}}

The accessible final states for non-relativistic
light dark fermions ($m_\chi\lesssim 100$ GeV) 
annihilating through the $U_Y(1)$ portal
\begin{equation}\label{eq:uy1}
\mathcal{L}_{B\chi}=\frac{1}{M_d}B_{\mu\nu}
\overline{\chi}S^{\mu\nu}(a_m+\mathrm{i}a_e\gamma_5)\chi
\end{equation} 
are pairs of fermions and anti-fermions ($f\overline{f}$)
and pairs of $U_Y(1)$ gauge bosons which in the low mass range yields $\gamma\gamma$.
However, the annihilation into the vector bosons is suppressed with $M_d^{-4}$.
The annihilation cross sections into $f\overline{f}$ states are for $s\ge 4m_f^2$
\begin{eqnarray}\nonumber
\sigma_{\chi\overline{\chi}\to f\overline{f}}&=&
N_c\frac{\alpha\cos^2\theta}{48M_d^2}\!\left(Y_{f,+}^2+Y_{f,-}^2\right)
\sqrt{\frac{s-4m_f^2}{s-4m_\chi^2}}
\\ \nonumber
&&\times\!\left(s-m_f^2\right)\!\left[a_m^2\!\left(s+8m_\chi^2\right)
+a_e^2\!\left(s-4m_\chi^2\right)\right]
\\ \nonumber
&&\times\!\left(\frac{1}{s^2}+\frac{2}{s}\frac{\left(s-m_Z^2\right)\tan^2\theta}{
\left(s-m_Z^2\right)^2+m_Z^2\Gamma_Z^2}\right.
\\ \label{eq:xx2ff}
&&+\left.\frac{\tan^4\theta}{
\left(s-m_Z^2\right)^2+m_Z^2\Gamma_Z^2}\right),
\end{eqnarray}
where $\alpha=e^2/4\pi$, $Y_{f,\pm}$ are the weak hypercharges of the right-
and left-handed fermions, respectively, and 
$N_c=1$ for leptons, $N_c=3$ for quarks.

In the light mass range of interest to us the $hZ$ and $hhZ$ final states
are not accessible in the non-relativistic regime and their contributions to
the thermally averaged cross section at thermal freeze-out are therefore negligible, 
but we also report the corresponding annihilation cross section into $hZ$ for completeness.
This cross section is with $s\ge (m_h+m_Z)^2$
\begin{eqnarray}\nonumber
\sigma_{\chi\overline{\chi}\to hZ}&=&2\pi\alpha^2\frac{\sin^2\theta}{\sin^4(2\theta)}
\frac{v_h^2}{M_d^2}
\\ \nonumber
&&\times\frac{a_m^2\!\left(s+8m_\chi^2\right)
+a_e^2\!\left(s-4m_\chi^2\right)}{\left(s-m_Z^2\right)^2+m_Z^2\Gamma_Z^2}
\\ \nonumber
&&\times\sqrt{\frac{s^2-2s\left(m_h^2+m_Z^2\right)+\left(m_h^2-m_Z^2\right)^2}{
s\!\left(s-4m_\chi^2\right)}}
\\ \label{eq:xx2hz}
&&\times\frac{s^2+2s\left(5m_Z^2-m_h^2\right)+\left(m_h^2-m_Z^2\right)^2}{12m_Z^2 s}.
\end{eqnarray}
This cross section becomes only relevant for masses $m_\chi\gtrsim 104$ GeV
(it is slightly lower than $(m_h+m_Z)/2$ due to the integration over $s$ in the thermal 
averaging), and the corresponding cross section into the $hhZ$ final state (which cannot 
be integrated analytically) becomes only relevant for masses $m_\chi\gtrsim 160$ GeV.

The cross sections determine the thermally averaged annihilation
cross section through the general formula from Gondolo and Gelmini \cite{gondolo}, 
\begin{eqnarray}\nonumber
\langle\sigma v\rangle(T)&=&
\int_{4m_\chi^2}^\infty\!\!ds\sqrt{s}\left(s-4m_\chi^2\right)\!\sigma(s)K_1(\sqrt{s}/T)
\\ \label{eq:thermalGG}
&&\times\frac{1}{8m_\chi^4 TK_2^2(m_\chi/T)}.
\end{eqnarray}
Note that (as usual) the low-energy effective cross sections for $s\ll M_d$ are 
sufficient for the freeze-out calculation since $K_1(\sqrt{s}/T)$ strongly 
suppresses all contributions for $s\gg 4m_\chi^2$. 

Spin averaging eliminated the cross-multiplication terms between the magnetic
and electric dipole couplings in the squares $\left|\mathcal{M}_{fi}\right|^2$
of the transition matrix elements for the annihilations. However, spin averaging 
also affects the magnetic and electric contributions very differently in such a way 
that 
\begin{equation}\label{eq:e2m}
\frac{\sigma(s)|_{a_e}}{\sigma(s)|_{a_m}}=\frac{a_e^2}{a_m^2}\frac{s-4m^2_\chi}{s+8m^2_\chi}.
\end{equation}
This implies that the low energy annihilation cross sections are dominated by the 
magnetic dipole coupling (unless $a_e^2\gg a_m^2$, which we do not assume). Therefore 
it is the magnetic dipole coupling $a_m/M_d\equiv 1/M$ which  determines the 
thermal freeze-out and the relic abundance of the electroweak dipole dark matter.

\begin{figure}[htb]
\begin{center}
\scalebox{0.9}{\includegraphics{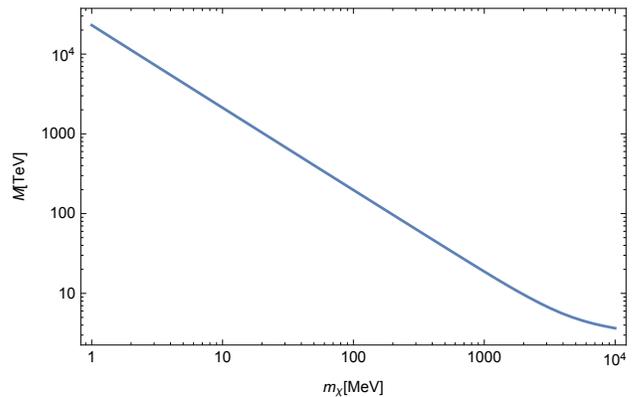}}
\caption{\label{fig:Mlowmass}
The magnetic dipole coupling scale $M=M_d/a_m$ as a function of dark matter 
mass $m_\chi$ in the range $1\,\mathrm{MeV}\le m_\chi\le 10\,\mathrm{GeV}$.}
\end{center}
\end{figure}

The thermally averaged cross section depends on the coupling scale $M$ through
the $M$-dependence of $\sigma(s)$,
\begin{equation}\label{eq:Mdep1}
\langle\sigma v\rangle(T)\equiv\langle\sigma v\rangle(M,T)
=\frac{1}{M^2[\mathrm{TeV}]}\langle\sigma v\rangle(1\mathrm{TeV},T),
\end{equation}
and the requirement for $\langle\sigma v\rangle(T)$ to match the required cross
section for thermal freeze-out \cite{LW,KT,dodelson} then determines the coupling
scale $M$ as a function of dark matter mass $m_\chi$. For $m_\chi\le 10$ GeV, $M$ decreases
with increasing $m_\chi$ with values $M\simeq 23$ EeV for $m_\chi=1$ MeV 
and $M\simeq 3.7$ TeV for $m_\chi=10$ GeV, see Fig. \ref{fig:Mlowmass}.

Since $M$ is related to the mass $m_C$ of a possible Kalb-Ramond field 
through $m_C=g_{BC}g_{C\chi}M$, coupling scales $M$ in the few TeV to thousands of TeV
range could indicate a Kalb-Ramond mass in the hundreds of GeV to hundreds of TeV range
if we assume weak strength couplings of the Kalb-Ramond field.

\section{Electron recoil cross section\label{sec:recoil}}

As explained in the white paper \cite{rouven1} on new ideas in dark matter research,
electron recoils are a primary possible signal for light dark matter particles
with electromagnetic dipole moments. 

The differential electron recoil cross section for the $U_Y(1)$ portal (\ref{eq:uy1}) 
in the lab frame and in the non-relativistic limit is for $m_e< m_\chi$ given by
\begin{eqnarray}\nonumber
\frac{d\sigma_e}{d\Omega}&=&\sum_{\pm}\frac{\alpha\cos^2\theta}{8\pi M^2}
\frac{m_e m_\chi^3}{(m_e+m_\chi)^2}\frac{1}{(\Delta\bm{k})_{\pm}^2}\!\left(
1+\frac{9a_e^2}{2a_m^2}\right)
\\  \label{eq:recoil1}
&&\times\frac{\left(\cos\varphi\pm\sqrt{(m_e/m_\chi)^2-\sin^2\varphi}\right)^2}{
\sqrt{(m_e/m_\chi)^2-\sin^2\varphi}},
\end{eqnarray}
where $\varphi$ is the scattering angle between the incoming and scattered
$\chi$ particle. The possible momentum transfers are in terms of the 
incoming $\chi$ momentum $\bm{k}$ and the scattering angle given by
\begin{eqnarray} \nonumber
\frac{(\Delta\bm{k})_{\pm}^2}{\bm{k}^2}&=&1
+\frac{\left(\cos\varphi\pm\sqrt{(m_e/m_\chi)^2-\sin^2\varphi}\right)^2}{
[1+(m_e/m_\chi)]^2}
\\ \label{eq:recoil2}
&&
-\,2\cos\varphi\frac{\cos\varphi\pm\sqrt{(m_e/m_\chi)^2-\sin^2\varphi}}{
1+(m_e/m_\chi)}.
\end{eqnarray}
The scattering angle is limited to 
\begin{equation}\label{eq:angle1}
\varphi\le\varphi_{max}=\arcsin(m_e/m_\chi).
\end{equation}
The two branches in the scattering cross section arise from the fact that for
$m_e<m_\chi$ there are two values of $|\bm{k}'|/|\bm{k}|$ for every scattering 
angle $\varphi<\varphi_{max}$. The $(+)$ branch corresponds to an 
increase from $\varphi=0$ to $\varphi=\varphi_{max}$ with the scattered
momentum $|\bm{k}'|$ decreasing from  $|\bm{k}'|=|\bm{k}|$ 
to $|\bm{k}'|=|\bm{k}|\sqrt{(m_\chi-m_e)/(m_\chi+m_e)}$.
The $(-)$ branch corresponds to a subsequent
decrease from $\varphi=\varphi_{max}$ to $\varphi=0$, with $|\bm{k}'|$ further
decreasing from $|\bm{k}'|=|\bm{k}|\sqrt{(m_\chi-m_e)/(m_\chi+m_e)}$
to $|\bm{k}'|=|\bm{k}|(m_\chi-m_e)/(m_\chi+m_e)$.

\begin{figure}[htb]
\begin{center}
\scalebox{0.65}{\includegraphics{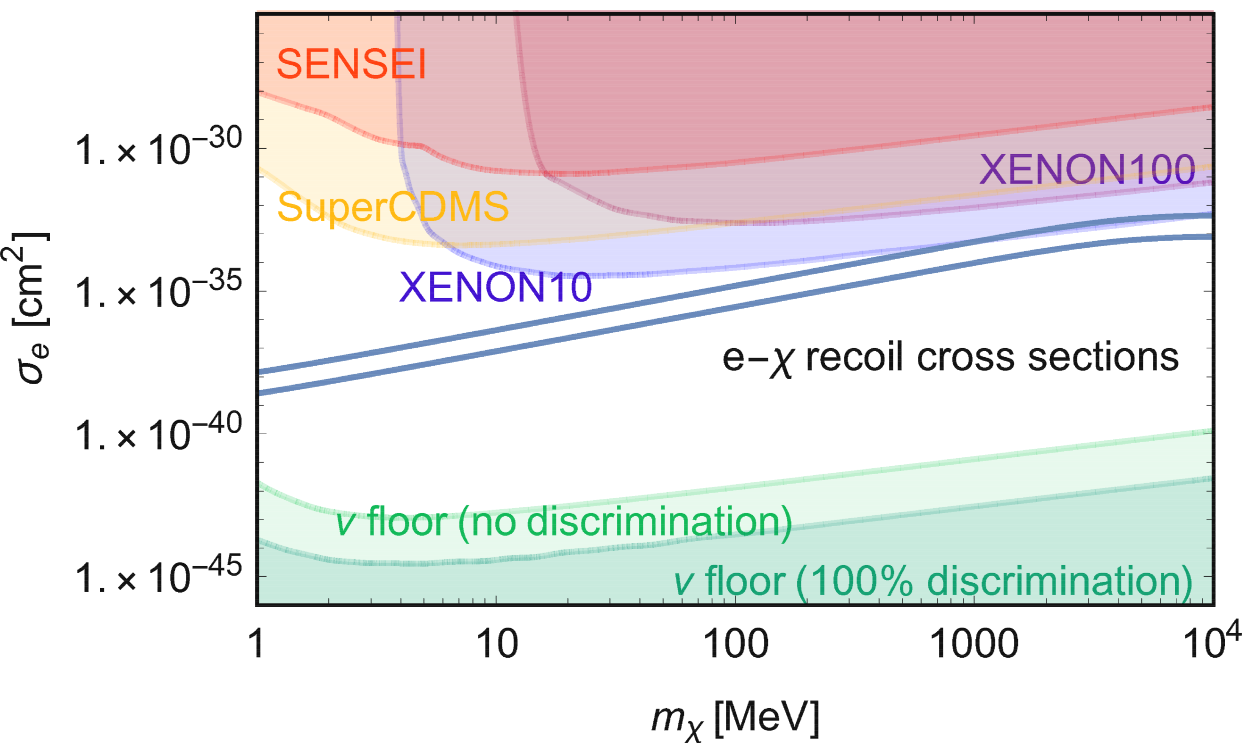}}
\caption{\label{fig:recoillowmass}
The electron recoil cross section as a function of dark matter 
mass $m_\chi$ in the range $1\,\mathrm{MeV}\le m_\chi\le 10\,\mathrm{GeV}$.
The lower curve is for $a_e=0$ and the upper curve is for $a_e^2=a_m^2$.
The neutrino floors with and without full separation of electron-neutrino
and nucleon-neutrino scattering are taken from Ref. \cite{floor}.}
\end{center}
\end{figure}

The dark matter abundance constraints in the previous section determine the magnetic 
coupling $M^{-1}=a_m/M_d$, but they do not determine $a_e/a_m$. Therefore we can
only calculate the recoil cross section as a function of dark matter mass if 
we assume a ratio $(a_e/a_m)^2$.
Integration of $d\sigma_{e}/d\Omega$ yields electron recoil cross 
sections which are below the current limits from SuperCDMS \cite{cdms}, 
XENON10/100 \cite{xenon}, and  SENSEI \cite{sensei}, if $a^2_e<a^2_m$. This is
displayed in Fig. \ref{fig:recoillowmass}, where
$\beta_\odot=8.47\times 10^{-4}$ was used as a fiducial dark matter speed \cite{pdg}.

The recoil cross sections from a magnetic dipole coupling comply with the current 
direct exclusion limits throughout the considered mass range. On the other hand, the 
case $a_e=\pm a_m$ is excluded for masses in the GeV range. We also note that the 
recoil cross sections are above the neutrino floor \cite{floor} and may be detectable 
with longer exposures or larger detectors.

\section{Conclusions\label{sec:conc}}

Gauge invariant interactions of open strings require the Kalb-Ramond field to couple
to the field strength tensors of $U(1)$ gauge fields, whereas a theory with only
closed strings permits Cremmer-Scherk couplings to $U(1)$ field strength tensors.
We found that these couplings can induce dipole couplings of electroweak singlet dark 
matter to the $U_Y(1)$ gauge field, thus contributing  to the formation
of a $U_Y(1)$ portal both to dark matter and to string theory. We analyzed in particular 
the case of a single dark matter component and found that the MeV-GeV mass range for 
dipole coupled dark matter remains viable 
under recent constraints from direct searches in electron recoils if $a_e^2<a_m^2$, 
while the case of dipole coupling only to right-handed or left-handed dark fermions
$a_e=\pm a_m$ is excluded in the GeV mass rang but still allowed in the MeV mass range.
Dipole coupled dark matter has a high 
discovery potential due to yielding recoil cross sections above the neutrino floor.
The discovery of a dipole coupled $U_Y(1)$ portal to dark matter would therefore be
very interesting from the perspective of a bottom-up approach to string phenomenology.

The model discussed here does not touch upon the important question of moduli 
stabilization, except for the observation that dilaton stabilization and an internal 
magnetic Kalb-Ramond flux decouple the Kalb-Ramond coupling in four dimensions from 
the mass term. We do assume that moduli are stabilized and that compactification yields 
the Standard Model at low energies. Our point is that under these circumstances the 
Kalb-Ramond field provides a natural candidate for inducing a dipole coupled $U_Y(1)$ 
portal to electroweak singlet dark matter. 
The discovery of $U_Y(1)$ dipole coupled dark matter would therefore provide an important 
low-energy indication for the existence of the anti-symmetric tensor fields of string theory.

\begin{acknowledgments} 
AD was supported by a MITACS Globalink internship. The research
of RD is supported by the Natural Sciences and Engineering Research 
Council of Canada (NSERC) through a subatomic physics grant. 
\end{acknowledgments}


\begin{thebibliography}{88}

\bibitem{ed}
M.W. Goodman, E. Witten, \emph{Detectability of certain dark matter candidates},
{Phys. Rev. D} {\bf 31} (1985) 3059.

\bibitem{frank}
B. Patt, F. Wilczek, \emph{Higgs-field portal into hidden sectors}, 
arXiv:hep-ph/0605188.

\bibitem{zee}
V. Silveira, A. Zee, \emph{Scalar Phantoms}, 
{Phys. Lett. B} {\bf 161} (1985) 136.

\bibitem{mcdonald}
J. McDonald, \emph{Gauge singlet scalars as cold dark matter}, 
{Phys. Rev. D} {\bf 50} (1994) 3637.

\bibitem{bento}
M.C. Bento, O. Bertolami, R. Rosenfeld, L. Teodoro, 
\emph{Selfinteracting dark matter and invisibly decaying Higgs},
{Phys. Rev. D} {\bf 62} (2000) 041302(R).

\bibitem{cliff}
C. Burgess, M. Pospelov, T. ter Veldhuis, 
\emph{The Minimal Model of nonbaryonic dark matter: a singlet scalar},
{Nucl. Phys. B} {\bf 619} (2001) 709.

\bibitem{dklm}
H. Davoudiasl, R. Kitano, T. Li, H. Murayama, \emph{The new Minimal Standard Model},
{Phys. Lett. B} {\bf 609} (2005) 117.

\bibitem{wells}
R. Schabinger, J.D. Wells, 
\emph{Minimal spontaneously broken hidden sector and its impact 
on Higgs boson physics at the CERN Large Hadron Collider},
{Phys. Rev. D} {\bf 72} (2005) 093007.

\bibitem{maxim}
C. Bird, R.V. Kowalewski, M. Pospelov, 
\emph{Dark matter pair-production in $b\to s$ transitions},
{Mod. Phys. Lett.} {\bf A 21} (2006) 457.

\bibitem{yann}
A. Djouadi, O. Lebedev, Y. Mambrini, J. Quevillon, 
\emph{Implications of LHC searches for Higgs-portal dark matter},
{Phys. Lett. B} {\bf 709} (2012) 65.

\bibitem{rd1}
R. Dick, \emph{Direct signals from electroweak singlets through the Higgs portal}, 
{Int. J. Mod. Phys. D} {\bf 27} (2018) 1830008.

\bibitem{gambit}
P. Athron \textit{et al.} (GAMBIT Collaboration),
\emph{Global analyses of Higgs portal singlet dark matter models using GAMBIT}, 
{Eur. Phys. J. C} {\bf 79} (2019) 38.

\bibitem{kris1} 
K. Sigurdson, M. Doran, A. Kurylov, R.R. Caldwell,
M. Kamionkowski, \emph{Dark-matter electric and magnetic dipole moments},
{Phys. Rev. D} {\bf 70} (2004) 083501.

\bibitem{masso}  
E. Masso, S. Mohanty, S. Rao, 
\emph{Dipolar dark matter},
{Phys. Rev. D} {\bf 80} (2009) 036009.

\bibitem{barger}  
V. Barger, W.-Y. Keung, D. Marfatia, 
\emph{Electromagnetic properties of dark matter: Dipole moments and charge form factor},
{Phys. Lett. B} {\bf 696} (2011) 74.

\bibitem{heo1} 
J.H. Heo, \emph{Electric dipole moment of Dirac fermionic dark matter}, 
{Phys. Lett. B} {\bf 702} (2011) 205.

\bibitem{tait} 
J.-F. Fortin, T.M.P. Tait,
\emph{Collider constraints on dipole-interacting dark matter},
{Phys. Rev. D} {\bf 85} (2012) 063506.

\bibitem{cline} 
J.M. Cline, G.D. Moore, A.R. Frey, 
\emph{Composite magnetic dark matter and the 130 GeV line},
{Phys. Rev. D} {\bf 86} (2012) 115013.

\bibitem{heo2}  
J.H. Heo, C.S. Kim, 
\emph{Dipole-interacting fermionic dark matter in positron, 
antiproton, and gamma-ray channels},
{Phys. Rev. D} {\bf 87} (2013) 013007.

\bibitem{PS}
S. Profumo, K. Sigurdson, 
\emph{Shadow of dark matter},
{Phys. Rev. D} {\bf 75} (2007) 023521.

\bibitem{also1}
A.L. Fitzpatrick, K.M. Zurek, \emph{Dark moments and the DAMA-CoGeNT puzzle},
{Phys. Rev. D} {\bf 82} (2010) 075004.

\bibitem{banks}
T. Banks, J.-F. Fortin, S. Thomas, 
\emph{Direct Detection of Dark Matter Electromagnetic Dipole Moments},
arXiv:1007.5515 [hep-ph].

\bibitem{also2}
B. Feldstein, P.W. Graham, S. Rajendran, \emph{Luminous dark matter},
{Phys. Rev. D} {\bf 82} (2010) 075019.

\bibitem{also3}
S. Chang, N. Weiner, I. Yavin, \emph{Magnetic inelastic dark matter},
{Phys. Rev. D} {\bf 82} (2010) 125011.


\bibitem{conlon}
J.P. Conlon, F. Day, N. Jennings, S. Krippendorf, M. Rummel, 
\emph{Consistency of Hitomi, XMM-Newton, and Chandra 3.5 keV data from Perseus}, 
{Phys. Rev. D} {\bf 96} (2017) 123009.

\bibitem{rouven1}
M. Battaglieri \textit{et al.}, 
\emph{US Cosmic Visions: New Ideas in Dark Matter 2017: Community Report},
arXiv:1707.04591 [hep-ph].

\bibitem{GSW}
M.B. Green, J.H. Schwarz, E. Witten,
\textit{Superstring Theory} Vols. 1 \& 2, Cambridge University Press, Cambridge (1987).

\bibitem{BLT}
R. Blumenhagen, D. L\"ust, S. Theisen,
\textit{Basic Concepts of String Theory}, Springer-Verlag, Berlin (2013).

\bibitem{KR0}
M. Kalb, P. Ramond, \emph{Classical direct interstring action},
{Phys. Rev. D} {\bf 9} (1974) 2273.

\bibitem{KR1}
A. Chatterjee, P. Majumdar, 
\emph{Kalb-Ramond field interactions in a braneworld scenario},
{Phys. Rev. D} {\bf 72} (2005) 066013.

\bibitem{KR2}
J. Erdmenger, R. Meyer, J.P. Shock, 
\emph{AdS/CFT with flavour in electric and magnetic Kalb-Ramond fields},
{JHEP} {\bf 0712} (2007) 091.

\bibitem{KR3}
 F.A. Barone, F.E. Barone, J.A. Helay\"{e}l-Neto, 
\emph{Charged branes interactions via Kalb-Ramond field},
{Phys. Rev. D} {\bf 84} (2011) 065026.

\bibitem{KR4}
Yun-Zhi Du, Li Zhao, Yi Zhong, Chun-E Fu, Heng Guo, 
\emph{Resonances of Kalb-Ramond field on symmetric and asymmetric thick branes},
{Phys. Rev. D} {\bf 88} (2013) 024009.

\bibitem{KR5}
I.C. Jardim, G. Alencar, R.R. Landim, R.N. Costa Filho,
\emph{Massive $p$-form trapping as a $p$-form on a brane},
{JHEP} {\bf 1504} (2015) 003.

\bibitem{KR6}
G. Alencar, I.C. Jardim, R.R. Landim,
\emph{$p$-Forms non-minimally coupled to gravity in Randall-Sundrum scenarios},
{Eur. Phys. J. C} {\bf 78} (2018) 367.


\bibitem{conf}
R. Dick, \emph{Conformal gauge fixing in Minkowski space}, 
{Lett. Math. Phys.} {\bf 18} (1989) 67.


\bibitem{modh1}
J. Polchinski, A. Strominger,
\emph{New vacua for type II string theory},
{Phys. Lett. B} {\bf 388} (1996) 736.

\bibitem{modh2}
J. Michelson,
\emph{Compactifications of type IIB strings to four dimensions with non-trivial
classical potential},
{Nucl. Phys. B} {\bf 495} (1997) 127.

\bibitem{modh3}
T.R. Taylor, C. Vafa,
\emph{RR flux on Calabi-Yau and partial supersymmetry breaking},
{Phys. Lett. B} {\bf 474} (2000) 130.

\bibitem{modh4}
G. Curio, A. Klemm, D. L\"ust, S. Theisen,
\emph{On the vacuum structure of type-II string compactifications on Calabi-Yau spaces
with H fluxes}, {Nucl. Phys. B} {\bf 609} (2001) 3.

\bibitem{modh5}
A.R. Frey, J. Polchinski,
\emph{$N=3$ warped compactifications},
{Phys. Rev. D} {\bf 65} (2002) 126009.

\bibitem{mod1}
S. Kachru, R. Kallosh, A.D. Linde, S.P. Trivedi,
\emph{de Sitter vacua in string theory},
{Phys. Rev. D} {\bf 68} (2003) 046005.

\bibitem{mod1b}
Yeuk-Kwan E. Cheung, S. Watson, R. Brandenberger,
\emph{Moduli stabilization with string gas and fluxes},
{JHEP} {\bf 0605} (2006) 025.

\bibitem{mod2}
I. Antoniadis, A. Kumar, T. Maillard,
\emph{Magnetic fluxes and moduli stabilization},
{Nucl. Phys. B} {\bf 767} (2007) 139.

\bibitem{mod3}
V. Balasubramanian, P. Berglund, J.P. Conlon, F. Quevedo,
\emph{Systematics of moduli stabilization in Calabi-Yau flux compactifications},
{JHEP} {\bf 0503} (2005) 007.

\bibitem{mod4}
M.R. Douglas, S. Kachru,
\emph{Flux compactification}, 
{Rev. Mod. Phys.} {\bf 79} (2007) 733.

\bibitem{mod5}
I. Antoniadis, J.-P. Derendinger, T. Maillard,
\emph{Nonlinear $N=2$ supersymmetry, effective actions and moduli stabilization},
{Nucl. Phys. B} {\bf 808} (2009) 53.

\bibitem{mod6}
B.S. Acharya, G. Kane, P. Kumar,
\emph{Compactified string theories - generic predictions for
particle physics},
{Int. J. Mod. Phys.} {\bf A 27} (2012) 1230012.

\bibitem{mod7}
 F. Quevedo,
\emph{Local string models and moduli stabilization},
{Mod. Phys. Lett. A} {\bf 30} (2015) 1530004.

\bibitem{mod8}
R. Blumenhagen, A. Font, M. Fuchs, D. Herschmann,
E. Plauschinn, Y. Sekiguchi, F. Wolf,
\emph{A flux-scaling scenario for high-scale moduli
stabilization in string theory},
{Nucl. Phys. B} {\bf 897} (2015) 500.

\bibitem{mod9}
I. Antoniadis, Y. Chen, G.K. Leontaris,
\emph{Perturbative moduli stabilization in type IIB/F-theory framework},
{Eur. Phys. J. C} {\bf 78} (2018) 766.

\bibitem{mod10}
W. Buchmuller, M. Dierigl, E. Dudas,
\emph{Flux compactifications and naturalness},
{JHEP} {\bf 1808} (2018) 151.


\bibitem{mtf1}
L.V. Avdeev, M.V. Chizhov, 
\emph{Antisymmetric tensor matter fields: An abelian model},
{Phys. Lett. B} {\bf 321} (1994) 212.

\bibitem{mtf2}
V. Lemes, R. Renan, S.P. Sorella,
\emph{Algebraic renormalization of antisymmetric tensor matter fields},
{Phys. Lett. B} {\bf 344} (1995) 158.

\bibitem{mtf3}
V. Lemes, R. Renan, S.P. Sorella,
\emph{Renormalization of nonabelian gauge theories with tensor matter fields},
{Phys. Lett. B} {\bf 392} (1997) 106.

\bibitem{CS}
E. Cremmer, J. Scherk, \emph{Spontaneous dynamical breaking of gauge symmetry in dual models}, 
{Nucl. Phys. B} {\bf 72} (1974) 117.

\bibitem{gondolo}
P. Gondolo, G. Gelmini, \emph{Cosmic abundances of stable particles: Improved analysis}, 
{Nucl. Phys. B} {\bf 360} (1991) 145.

\bibitem{LW}
B.W. Lee, S. Weinberg, \emph{Cosmological lower bound on heavy-neutrino masses},
{Phys. Rev. Lett.} {\bf 39} (1977) 165.

\bibitem{KT}
E.W. Kolb, M.S. Turner,
\emph{The Early Universe}, Westview Press, Boulder (1994).

\bibitem{dodelson}
S. Dodelson, \emph{Modern Cosmology}, Academic Press, San Diego (2003).

\bibitem{cdms}
R. Agnese \textit{et al.} (SuperCDMS Collaboration), 
\emph{First dark matter constraints from a SuperCDMS single-charge sensitive detector},
{Phys. Rev. Lett.} {\bf 121} (2018) 051301.

\bibitem{xenon}
R. Essig, T. Volansky, T.-T. Wu, 
\emph{New constraints and prospects for sub-GeV dark matter scattering off electrons in xenon},
{Phys. Rev. D} {\bf 96} (2017) 043017.

\bibitem{sensei}
M. Crisler \textit{et al.} (SENSEI Collaboration), 
\emph{SENSEI: First direct-detection constraints on sub-GeV dark matter from a surface run},
{Phys. Rev. Lett.} {\bf 121} (2018) 061803.

\bibitem{pdg}
M. Tanabashi \textit{et al.} (Particle Data Group), \emph{Review of Particle Physics}, 
{Phys. Rev. D} {\bf 98} (2018) 030001. 

\bibitem{floor}
J. Wyenberg, I.M. Shoemaker, 
\emph{Mapping the neutrino floor for direct detection experiments 
based on dark matter-electron scattering},
{Phys. Rev. D} {\bf 97} (2018) 115026.

\end{thebibliography}
\end{document}